\documentclass[onecolumn,showpacs,preprintnumbers,amsmath,amssymb,pra,epsf]{revtex4}
\usepackage{graphicx}

\newcommand{\bea}{\vspace{0.25cm}\begin{eqnarray}}
\newcommand{\eea}{\end{eqnarray}}

\begin{document}
\title{Go and return propagation of biphotons in fiber and polarization entanglement}

\author{G. Brida, M. Genovese, L.~A.~Krivitsky}
\address{Istituto Nazionale di Ricerca Metrologica,
Strada delle Cacce 91, 10135 Torino, Italy }
\author{M.~V.~Chekhova }
\address{Department of Physics, M.V.Lomonosov Moscow State
University,  Leninskie Gory, 119992 Moscow, Russia}
\author{E. Predazzi }
\address{Dip. Fisica Teorica
Univ. Torino and INFN, via P. Giuria 1, 10125 Torino, Italy}

%\ead{genovese@inrim.it}
\begin{abstract}

Propagation of entangled photons in optical fiber is one of the
fundamental issues for realizing quantum communication protocols.
When entanglement in polarization is considered, arises the problem
of compensating for the fiber effect on photons polarization. In
this paper we demonstrate an effective solution where a Faraday
mirror allows to cancel undesired effects of polarization drift in
fiber. This technique is applied to a protocol for generating Bell
states by a narrow temporal selection of the second-order intensity
correlation function.

\end{abstract}


\maketitle

\section{Introduction}

Transmission of photons in optical fibers is a fundamental tool both
for realizing quantum communication protocols \cite{gis} and for
experiments addressed to test foundations of quantum mechanics
\cite{pr}. For example, long-distance quantum key distribution
protocols in fiber have been realized by exploiting interferometric
schemes \cite{long} up to 100 km.

When quantum communication protocols are realized with photon
states encoded in polarization some general difficulty must be
considered connected with the transmission of photons through
optical fibers. The photons propagating in fibers experience
random polarization transformations due to the fluctuations of the
fiber birefringence. These fluctuations are caused by mechanical,
acoustic and thermal stresses in the fiber. Since it is impossible
to completely isolate the fiber from the external disturbance the
problem of an effective compensation for the polarization drift
becomes a crucial issue. For instance a Quantum Key Distribution
(QKD) protocol with polarization-entangled states produced via a
type-II parametric down conversion (PDC) source was realized
between a bank and the Vienna city hall by using a 1.45 km optical
fiber \cite{zeicryp}. Another example is a QKD network, involving
also fiber propagation of polarization entangled photons,  built
among several research institutes in Boston \cite{darpa}. In these
examples a correction for the fiber effect on polarization was
introduced, moreover, this correction was tuned in time quite
often due to the instability of the fiber. A similar correction
has also been needed in experiments on the study of the
interference structure in the second-order intensity correlation
function \cite{nosfib,nosfib2}, visualized due to the spreading of
the two-photon wave packet propagating through an optical fibre.

However, in experiments where the photon travels along the fiber in
both directions (there and back), a more efficient scheme can be
adopted. Indeed in \cite{mart1} it was shown that a Faraday cell
followed by a mirror acts as a temporal inversion matrix, $
\left[\begin{array}{cc}
0 & -1 \\
-1 & 0
\end{array}\right]
$ in the Jones formalism. Thus the effect of the fiber on the
polarization is reversed and cancels out when the photons are
reflected back to the fiber by a Faraday mirror. Let us notice
that a similar transformation cannot be obtained with retardation
plates and mirrors. An experiment with polarized light proved the
viability of the  scheme \cite{mart2}, which was later
successfully applied to realize an interferometer \cite{gisint}.
In particular, such configuration can find a very useful
application in "go and return" quantum communication protocols as
the ones of Ref. \cite{gob}.

In this paper we discuss the application of the Faraday mirror to
observe quantum interference in the shape of the second-order
correlation function where bipartite polarization entangled states
are produced in a type-II PDC process.

\section{Theory}

We investigate degenerate type-II PDC emission in the collinear
direction to the pump beam (the photons of a pair have the same
frequencies and orthogonal polarizations). In low gain regime,
neglecting higher orders, the state vector of the  PDC light can
be represented as a superposition of the vacuum state and  the
two-photon state given by an integral over the  spectrum
 \bea
|\Psi\rangle=|vac\rangle+\int{d\Omega}F(\Omega)[a^{\dagger}_{H}(\omega_0+\Omega)
a^{\dagger}_{V}(\omega_0-\Omega)e^{i\Omega\tau_0} \nonumber + \nonumber \\
a^{\dagger}_{V}(\omega_0+\Omega)
a^{\dagger}_{H}(\omega_0-\Omega)e^{-i\Omega\tau_0}]|vac\rangle,
\label{mainstate} \eea
 with $\omega_0=\omega_p/2$, $\omega_p$ being the pump
frequency, and $a^{\dagger}_{H}$ and  $a^{\dagger}_{V}$ being the
photon creation operators in the horizontal and vertical
polarization modes (denoted by $H,V$). The phase factor
$e^{\pm{i\Omega\tau_0}}$ is defined by the average temporal delay
between orthogonally polarized photons, where $\tau_0={DL/2}$,
$D\equiv{1/u_V-1/u_H}$ being the difference of the inverse group
velocities and $L$ the length of the crystal. The spectral and
temporal properties of two-photon light are described by the
two-photon spectral amplitude $F(\Omega)$ which establishes the
natural bandwidth of PDC. The second-order correlation function
$G^{(2)}(\tau)$ (the observable typically measured in experiments
with PDC) is given by the square modulus of the Fourier transform
of $F(\Omega)$, which is called the biphoton amplitude
\cite{Rubin},
\begin{equation}
G^{(2)}(\tau)\propto|F(\tau)|^2=|\int{d\Omega}F(\Omega)\exp(i\Omega\tau)|^2.
\label{gtwo}
\end{equation}

When the state (\ref{mainstate}) is transmitted through an optical
fiber with Group-Velocity Dispersion (GVD), as it has been shown
in \cite{valencia}, a Fourier transformation is performed over the
biphoton amplitude $F(\tau)$. Thus, this takes the shape of the
two-photon spectral amplitude $F(\Omega)$. This effect has a clear
analogue with the propagation of a short optical pulse in a GVD
media and results in changing the shape of the pulse to the shape
of its spectrum. As it was shown in \cite{nosfib}, $G^{(2)}(\tau)$
shows an interference structure depending on the polarization
selection performed over each photon of a pair. Namely, if after
dividing the  two photons on a 50/50 beamsplitter \cite{twoarms},
they are registered either with the same orientations of
polarization filters set at $45^\circ$ to initial basis or with
the orthogonal orientations of the polarization filters set at
$45^\circ$ and $-45^\circ$, $G^{(2)}(\tau)$ takes the form

\begin{eqnarray}
G_{+}^{(2)}(\tau)\sim\frac{\sin^2(\tau/\tau_f)\cos^2(\tau/\tau_f)}{(\tau/\tau_f)^2},\nonumber\\
G_{-}^{(2)}(\tau)\sim\frac{\sin^4(\tau/\tau_f)}{(\tau/\tau_f)^2}.
\label{sinecosine}
\end{eqnarray}

In Eq.\ref{sinecosine} $G_{+}^{(2)}$ and $G_{-}^{(2)}$ correspond,
respectively, to parallel (both at $45^\circ$) and orthogonal (one
at $45^\circ$ and another one at $-45^\circ$) orientations of the
polarization filters in the output ports of the beamsplitter.
$\tau_f\equiv2k''z/\tau_0$ is the typical width of the correlation
function after the fibre ~\cite{Krivitsky},  $z$ being the length
of the crystal and $k''$ the second derivative of the fibre
dispersion law $k(\omega)$.

Now let us focus on the problem of polarization drift in the fiber
and its influence on the polarization entanglement. Upon passing
the fiber, the initial polarization state of each photon
(horizontal or vertical) is transformed from the equator of the
Poincar\'e sphere to some arbitrary point on its surface
\cite{tomo}. This unitary transformation can be modelled as a
polarization rotation by means of a retardation plate with an
arbitrary optical thickness $\delta$ and orientation of the
optical axis to initial basis $\alpha$ \cite{delta}. Applying the
Jones formalism to the transformation of the creation operators in
(\ref{mainstate}) it can be shown that for a given parameters of
the retardation plate $\delta$ and $\alpha$, the shape of
$G_{+}^{(2)}$ and $G_{-}^{(2)}$ change  from (\ref{sinecosine}) to

\begin{eqnarray}
G_{+}^{(2)}(\tau)\sim[\cos^2\delta(1+\sin^2\delta)+\sin^4\delta\cos^2(4\alpha)]
\frac{\sin^2(\tau/\tau_f)\cos^2(\tau/\tau_f)}{(\tau/\tau_f)^2},\nonumber\\
G_{-}^{(2)}(\tau)\sim[\sin^2(4\alpha)\sin^4\delta\cos^2(\tau/\tau_f)+\sin^2(\tau/\tau_f)]
\frac{\sin^2(\tau/\tau_f)}{(\tau/\tau_f)^2}. \label{sinecosineplate}
\end{eqnarray}

The general dependence of $G_{\pm} $ on the parameters $\delta$
and $\alpha$ is shown in Fig.1. From this one can prove the
complexity of the  variation of the interference pattern as a
function of $\alpha$ and $\delta$.

Let us now focus on the generation of Bell states.
 As it was shown in \cite{nosfib} a Bell state can be produced by performing a narrow temporal
post-selection in the shape of $G^{(2)}(\tau)$. This can be
realized, for instance, by analyzing the $G^{(2)}(\tau)$
distribution with a Time to Amplitude Converter (TAC) and a Multi
Channel Analyzer (MCA), in which only specific channels are
selected. When one selects the central part of $G^{(2)}(\tau)$
corresponding to zero delay, i.e. $\tau = 0$ (which is equivalent
to the selection of $\Omega=0$ in the frequency domain), the state
(\ref{mainstate}) becomes the polarization-spatial entangled Bell
state: $|\psi^+\rangle=\{a^{\dagger}_{H1}a^{\dagger}_{V2}
+a^{\dagger}_{V1} a^{\dagger}_{H2}\}|vac\rangle,$ where
$a^{\dagger}_{\sigma i}$ are photon creation operators in the
horizontal and vertical polarization modes (denoted by
$\sigma=H,V$) and in two spatial modes of the beamsplitter
(denoted by $i=1,2$). Being a pure maximally entangled state,
$|\psi^+\rangle$ manifests $100\%$ visibility of polarization
interference \cite{visibility}. However, if we consider the
transformation of $|\psi^+\rangle$ in fiber, for instance modelled
for the sake of simplicity by a Half Wave Plate (HWP) with
variable orientation $\alpha$, then, as one can see from Fig.2 (a
section of Fig.1), the visibility of polarization interference at
$\tau = 0$ changes from $100\%$ for the case when $\alpha=0^\circ$
(which is equivalent to the absence of rotation) to zero  when
$\alpha=11,25^\circ$. Moreover, at $\alpha=22,5^\circ$ the
two-photon state becomes an eigen-state of the measurement basis
(photons of the pair are polarized at $45^\circ$ and $-45^\circ$)
and the interference structure is completely erased. Thus, even
considering the simplest model of polarization rotation produced
by the fiber, this leads to  spoiling the polarization
interference and to erasing the polarization entanglement.

This problem has motivated our choice of applying a Faraday mirror
to compensate for these effects. Moreover, taking into account a
wide natural bandwidth of PDC spectra (typically of the order of
ten nanometers) it is worth mentioning that the implementation of
an appropriate Faraday mirror allows also to correct for the
effects related to frequency dependence of the effective
birefringence in the fiber on the wavelength.

On the other hand, if the selected time interval corresponds to
$\tau=\pm\pi\tau_f/{2}$, the relative phase between the two
components of the state (\ref{mainstate}) becomes equal to $\pi$
and the biphoton state represents the polarization-frequency
singlet Bell state:
$|\psi^-\rangle=\{a^{\dagger}_{H1}(\omega_1)a^{\dagger}_{V2}(\omega_2)
-a^{\dagger}_{V1}(\omega_1)
a^{\dagger}_{H2}(\omega_2)\}|vac\rangle$, where the two frequency
modes are
$\omega_1=\omega_0+\pi/{2\tau_0};\omega_2=\omega_0-\pi/{2\tau_0}$.
As one can see from (\ref{sinecosineplate}), the visibility of
polarization interference in this case remains $100\%$ and does
not depend on the polarization rotation induced by the fiber. This
confirms a noticeable property of the singlet Bell state, namely,
its invariance with respect to any polarization transformation.
Therefore for the case of the selection of $|\psi^-\rangle$ no
compensation for the polarization drift is needed at all
\cite{kw}.

\section{Experimental set-up and data}

In our set-up, see Fig.3, biphoton pairs were generated via
spontaneous parametric down-conversion by pumping a type-II 0.5 mm
BBO crystal with a 0.5 Watt $\hbox{Ar}^{+}$ cw laser beam at the
wavelength 351 nm in the collinear frequency-degenerate regime.
After the crystal, the pump laser beam was eliminated by a 95\%
reflecting UV mirror and the PDC radiation was coupled into a 240
m long single-mode non polarization maintaining fibre with field
mode diameter $=4\mu$m by a 20x microscope objective lens placed
at the distance of 50 cm. This scheme provided imaging of the pump
beam waist onto the fibre input with the magnification 1:40. For
the efficient use of the pump beam, the pump was focused into the
crystal using a UV-lens with the focal length of 30 cm. The
resulting beam waist diameter, $120\mu m$, was small enough to be
coupled with the fibre core diameter, but still sufficiently large
not to influence the PDC angular spectrum.

After the first passage through the fiber, biphotons were
addressed back by a free-space Faraday mirror, consisting of a
Faraday rotator and a high-reflection mirror. In this way the
polarization effects of the fiber were compensated. Then, after
having crossed the crystal, the photons passed the UV mirror
needed for the injection of the pump beam into the crystal and
reached the 50/50 beam splitter preceding the detection
apparatuses, consisting of polarizers and two avalanche
photodiodes. The photocount pulses of the two detectors were sent
to the START and STOP inputs of a TAC. The output of the TAC was
finally addressed to a MCA, and the distribution of coincidences
over the time interval between the photocounts of two detectors
was observed at the MCA output.

The second-order intensity correlation function measured in the
experiment for the case when both polarizers are parallel (set at
$45^\circ$) and orthogonal (set at $45^\circ$ and $-45^\circ$) are
presented in Fig.4. The results show a good agreement with the
theoretical predictions (\ref{sinecosine}). Here we would like to
stress that the fiber was not placed in any isolation box or similar
device so that it experienced the influence of heating in the lab,
acoustic and mechanical stresses during all the acquisition time
which exceeded two hours. Moreover, the fiber we used was unwound
from the spool and did not have an isolation plastic core. We have
checked that in a single--pass configuration configuration under the
same conditions the polarization effects of the fiber drift on a
time scale of 5-7 minutes \footnote{The strong coupling between
fiber and environment and the polarization effects induced by this
are the most important problem for quantum communication with
polarization as qubit, on the other hand this coupling can be useful
for studying properties of quantum channels as Complete Positivity
\cite{BF} or decoherence control protocols.}.

Thus our results demonstrate that  a perfect stability of the
setup in "go and return" configuration over polarization
fluctuations introduced by the fibre can be achieved by using a
Faraday mirror.

The visibility of polarization interference at zero time delay
measured in the experiment and calculated from Fig.4 was $72\%$
(after
 background substraction). This reduction of the measured visibility
 with respect to theoretical predictions
is explained by the length of the fibre, insufficient to make the
spreading of $G^2(\tau)$  significantly  larger than the time
resolution of the set-up. Indeed, the fibre was chosen to be only
$2\times240$m long in order to avoid high losses of the signal
(about 12dB/km at 702 nm). A more detailed discussion on the
influence of the fiber length on the visibility of polarization
interference can be found in \cite{nosfib2}.

\section{Conclusions}

In conclusion we demonstrated an application of the "go and
return" scheme for generating entangled states with the help of
the spreading of the second-order intensity correlation function.
We have shown that the implementation of the Faraday mirror allows
to compensate for the polarization drift in the fiber, which is
crucial for obtaining polarization entangled Bell states. Our
experimental data verifies the stability of the scheme under the
influence of external disturbances over the fiber.

\section{Acknowledgements}
This work was supported by MIUR (FIRB RBAU01L5AZ-002 and PRIN
2005023443-002 ), by Regione Piemonte (E14), and by "San Paolo
foundation". Maria Chekhova also acknowledges the support of the
Russian Foundation for Basic Research, grant no.06-02-16393.

We would like also to acknowledge Luca Giacone for his assistance
in the realization of this experiment.

\begin{figure}
\begin{tabular}{cc}
\includegraphics[width=0.25\textwidth]{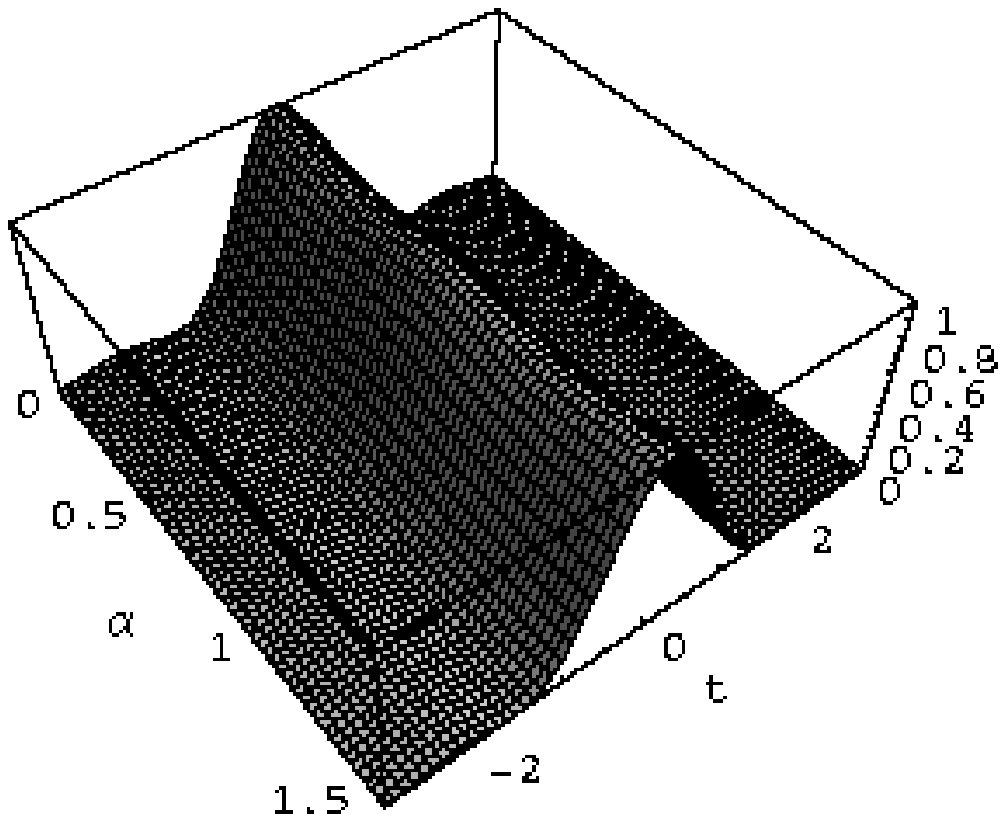}
\includegraphics[width=0.25\textwidth]{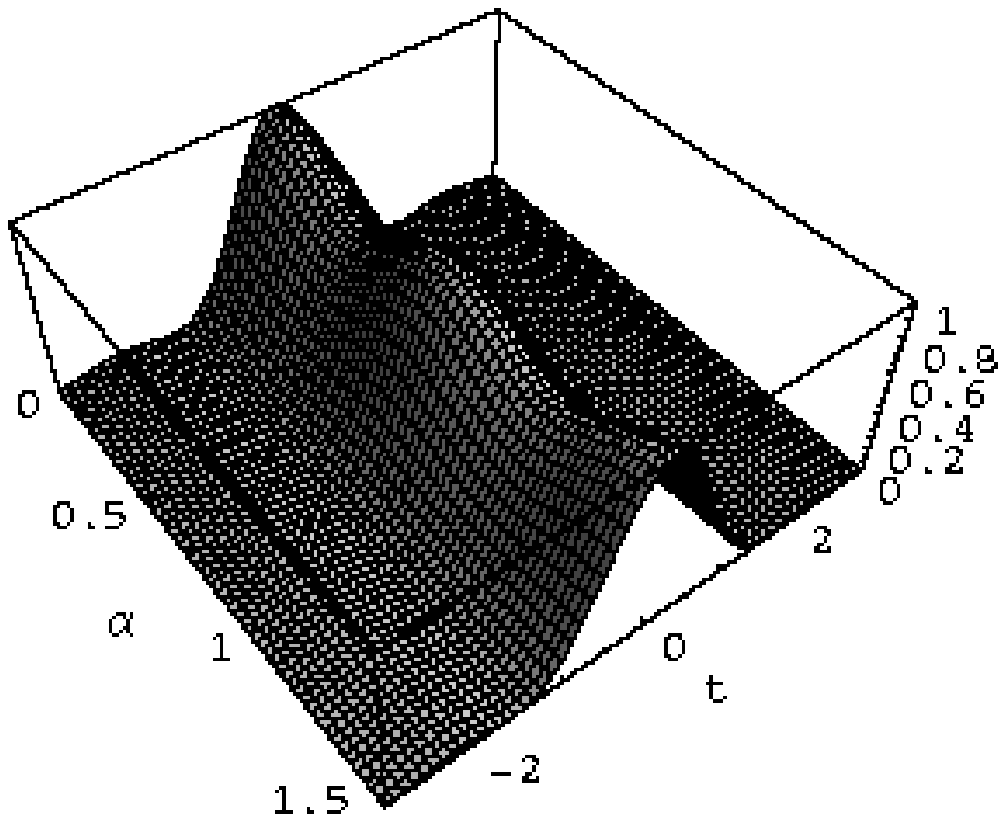}
\includegraphics[width=0.25\textwidth]{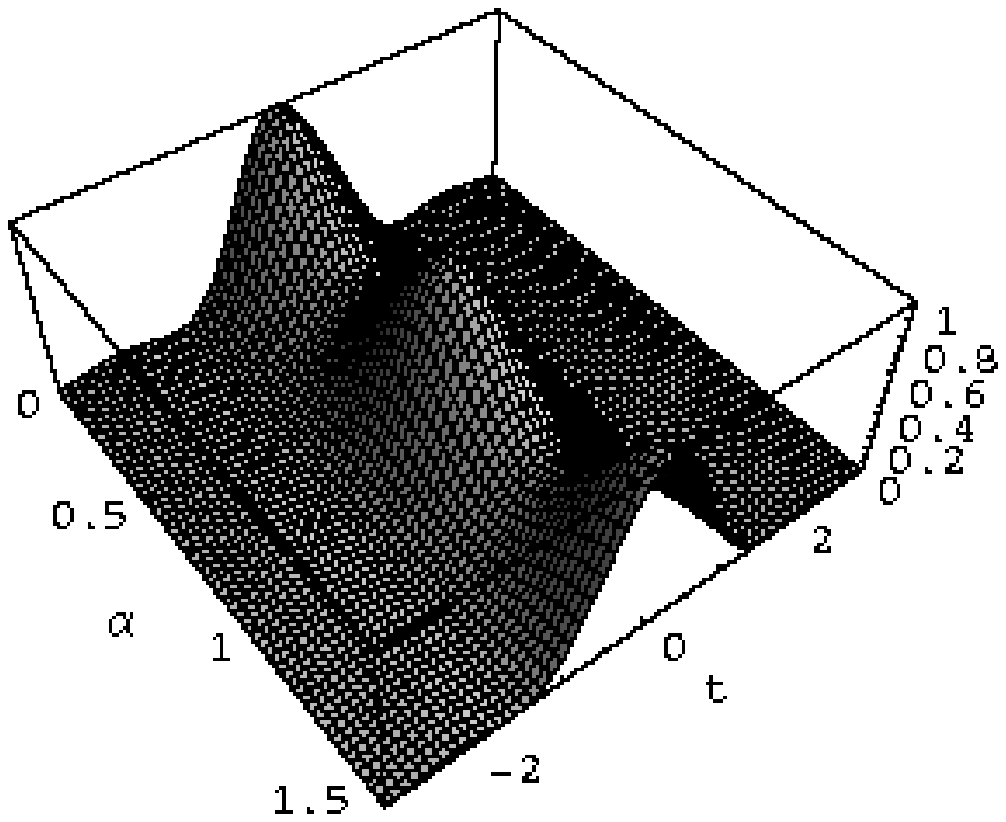}
 \includegraphics[width=0.25\textwidth]{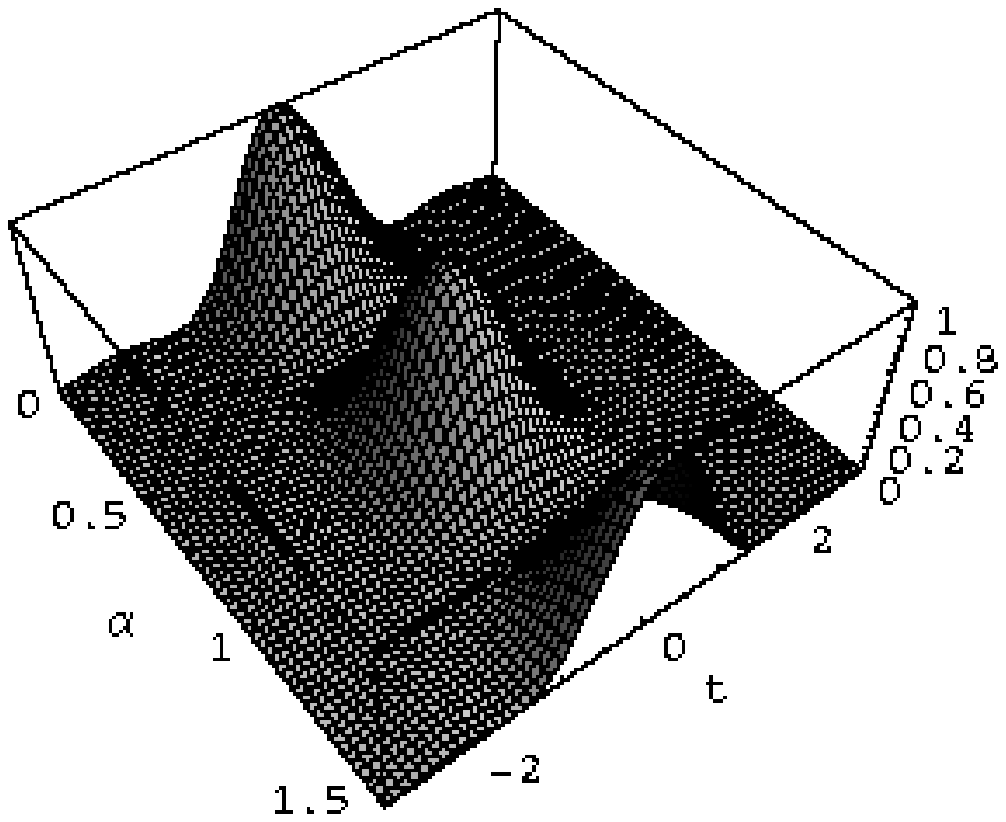} \cr
\includegraphics[width=0.25\textwidth]{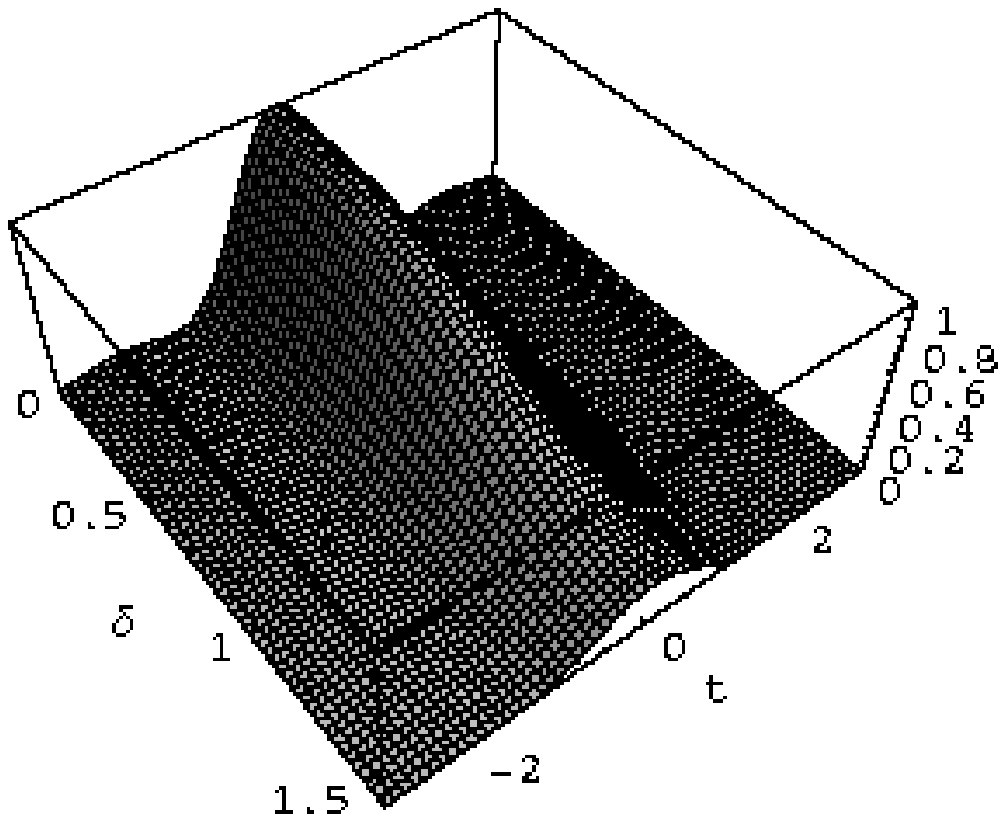}
\includegraphics[width=0.25\textwidth]{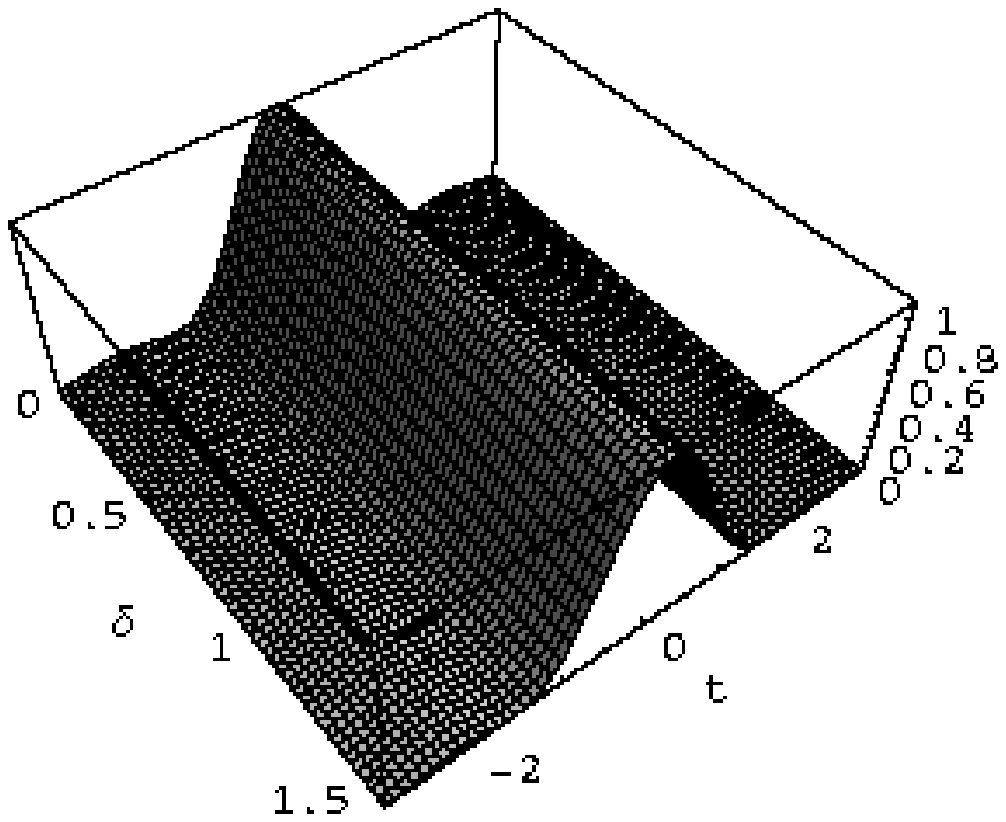}
\includegraphics[width=0.25\textwidth]{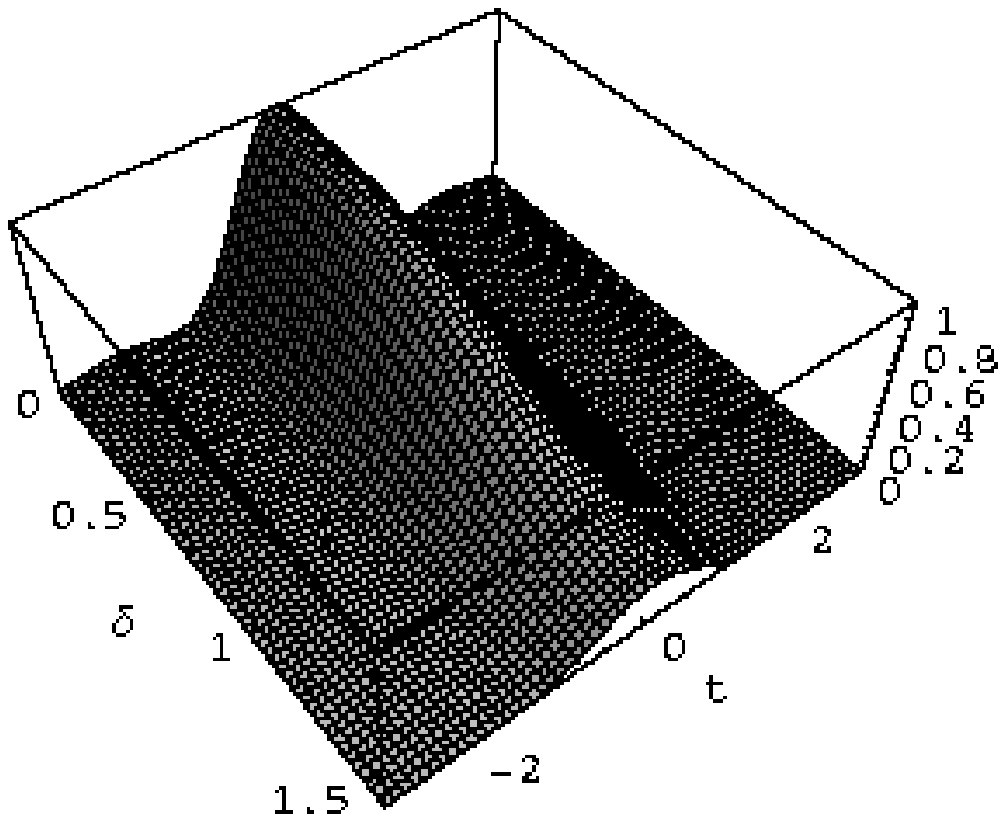}
 \includegraphics[width=0.25\textwidth]{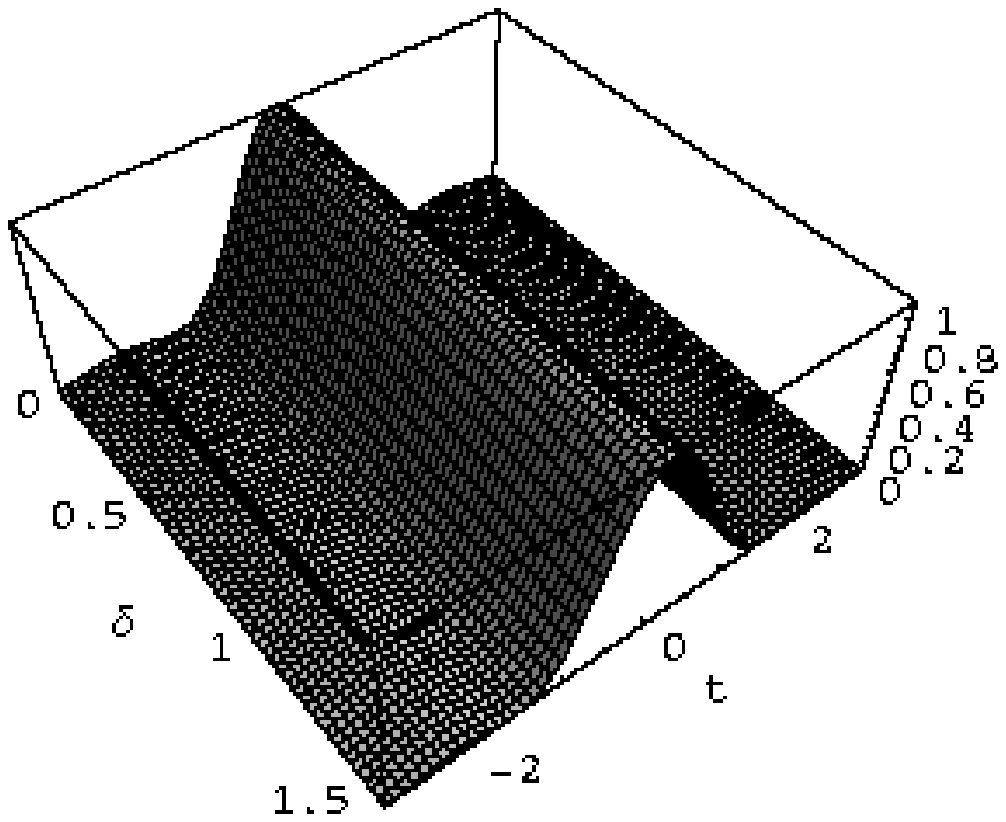} \cr
\includegraphics[width=0.25\textwidth]{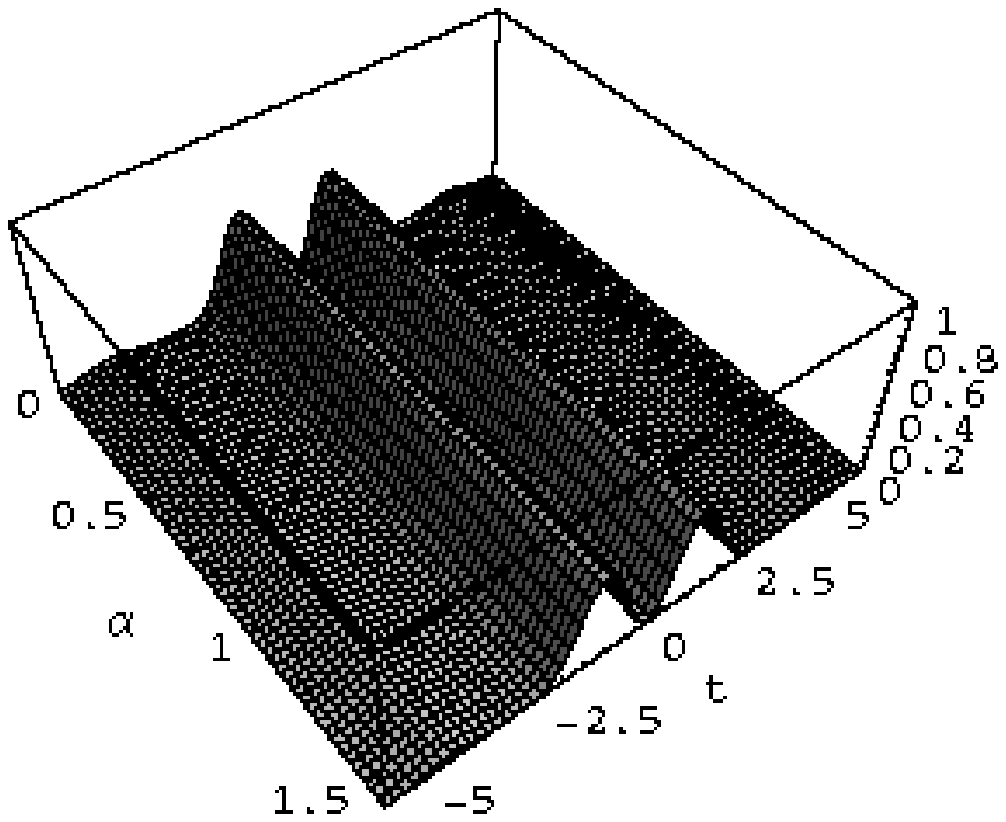}
\includegraphics[width=0.25\textwidth]{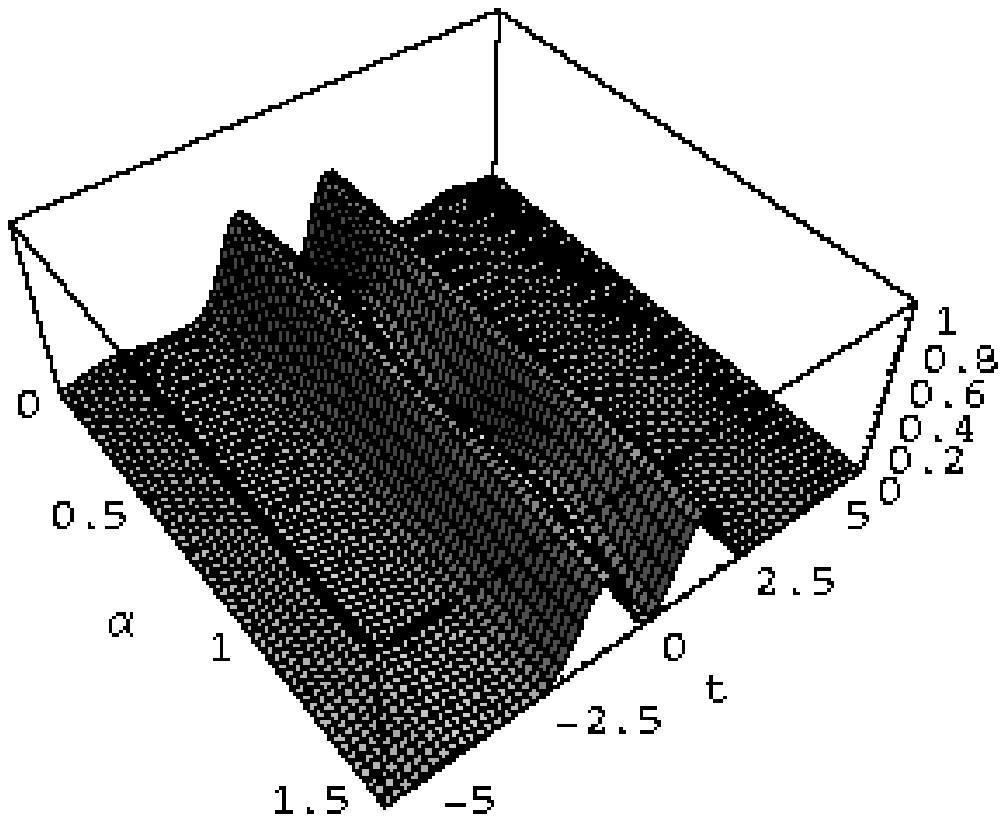}
\includegraphics[width=0.25\textwidth]{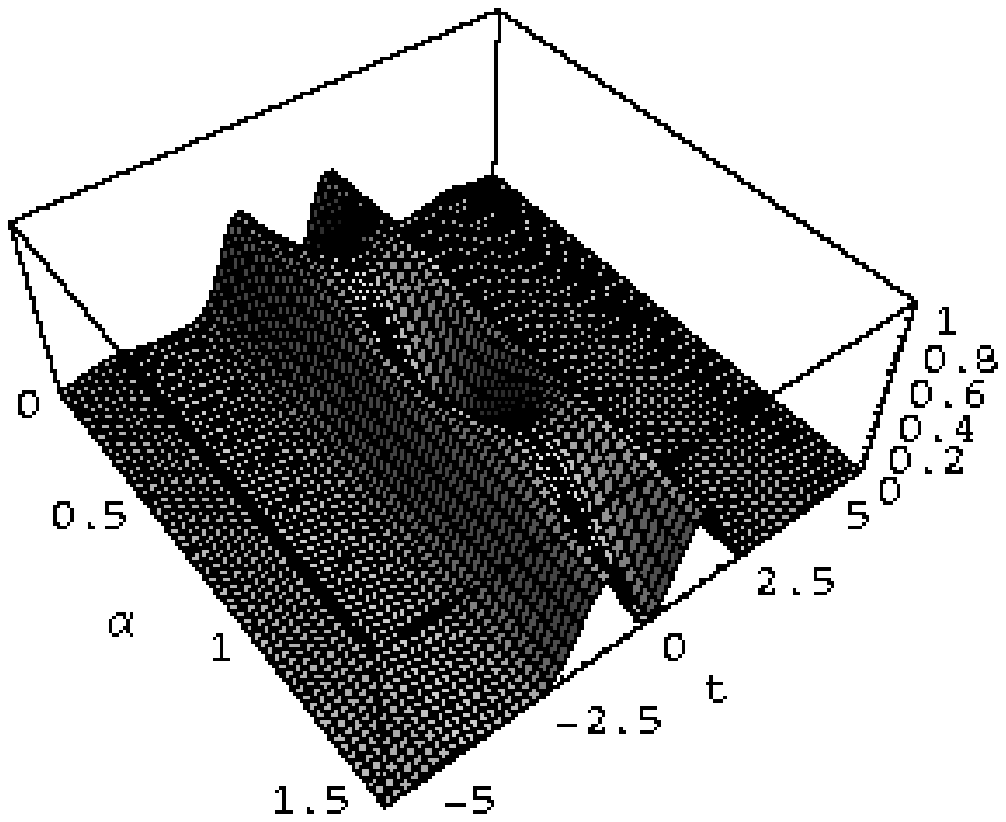}
 \includegraphics[width=0.25\textwidth]{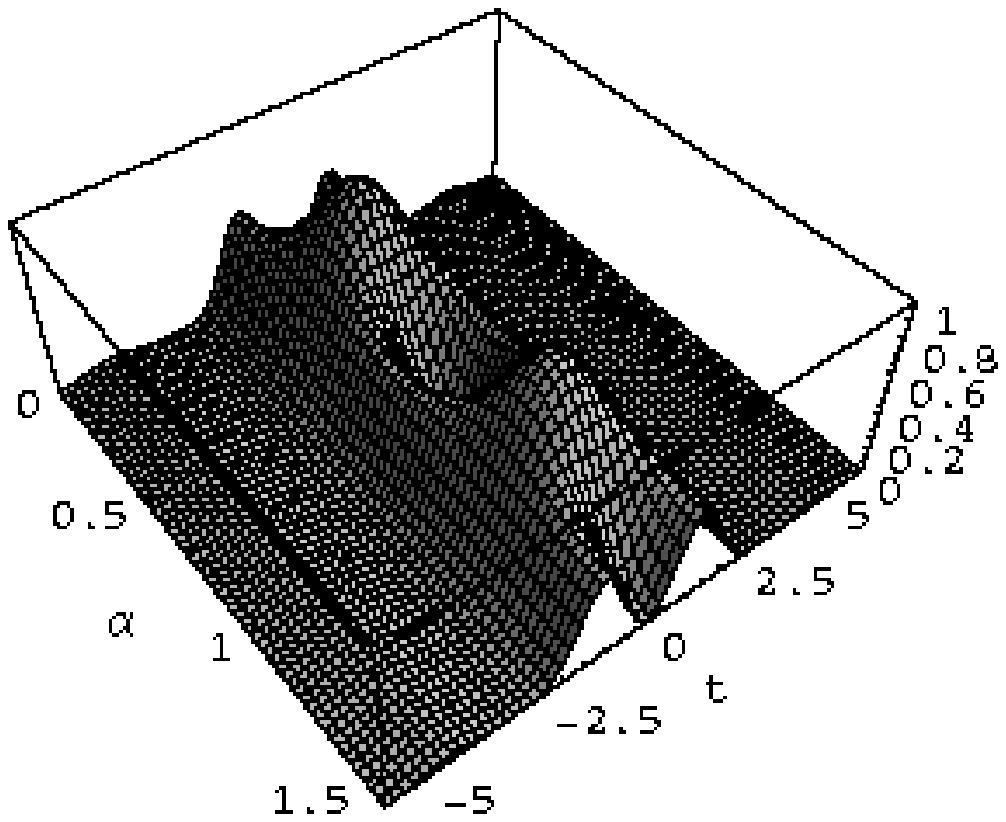} \cr
\includegraphics[width=0.25\textwidth]{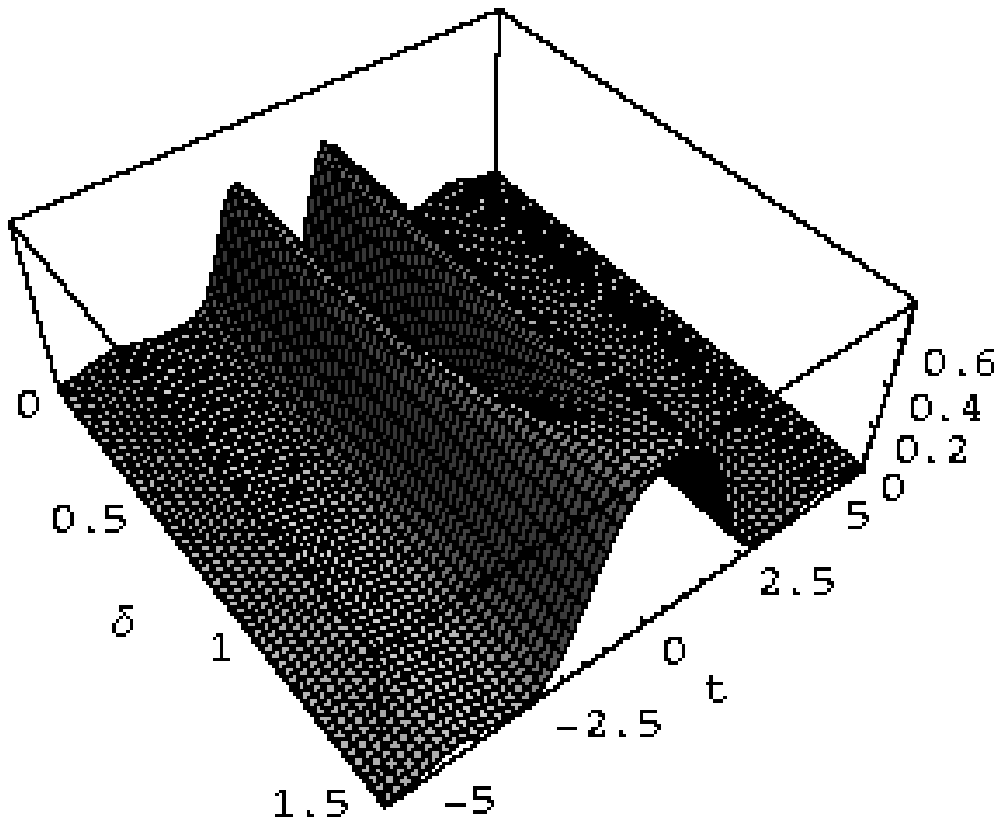}
\includegraphics[width=0.25\textwidth]{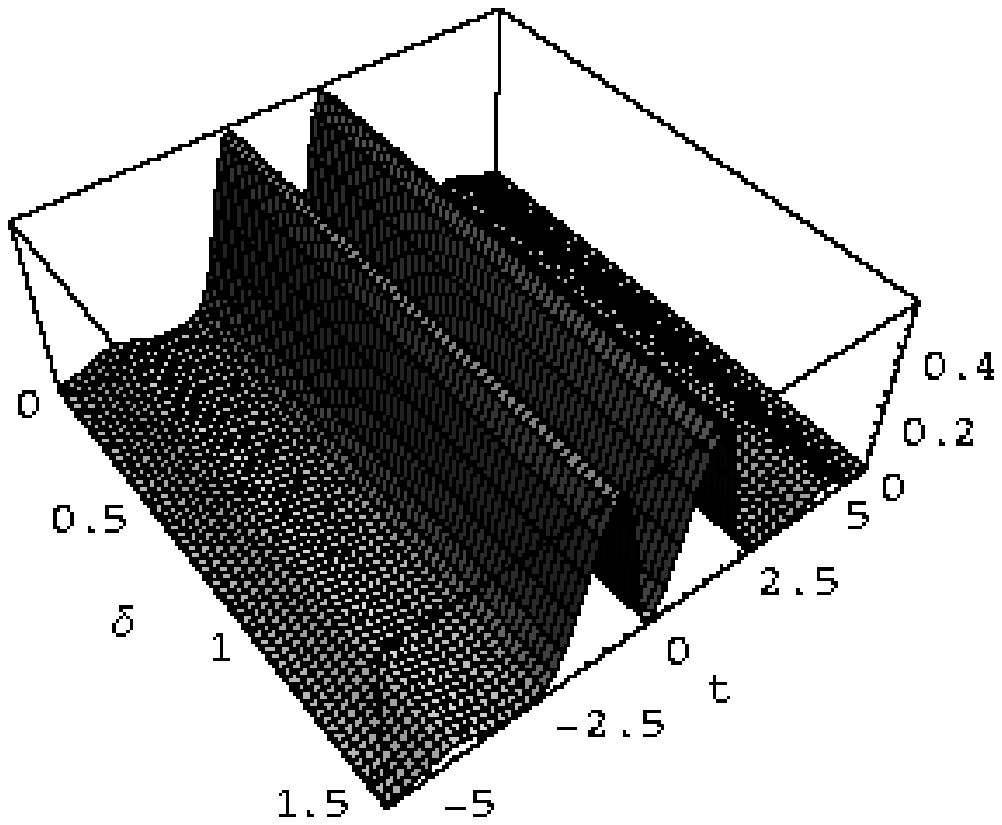}
\includegraphics[width=0.25\textwidth]{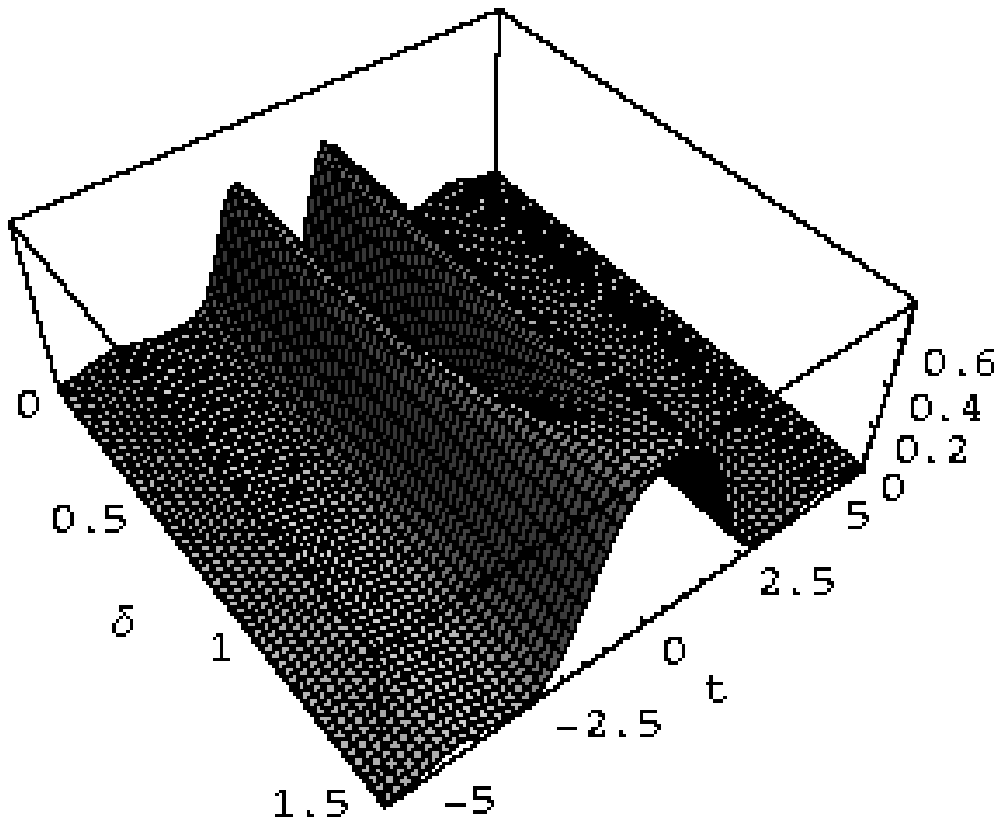}
 \includegraphics[width=0.25\textwidth]{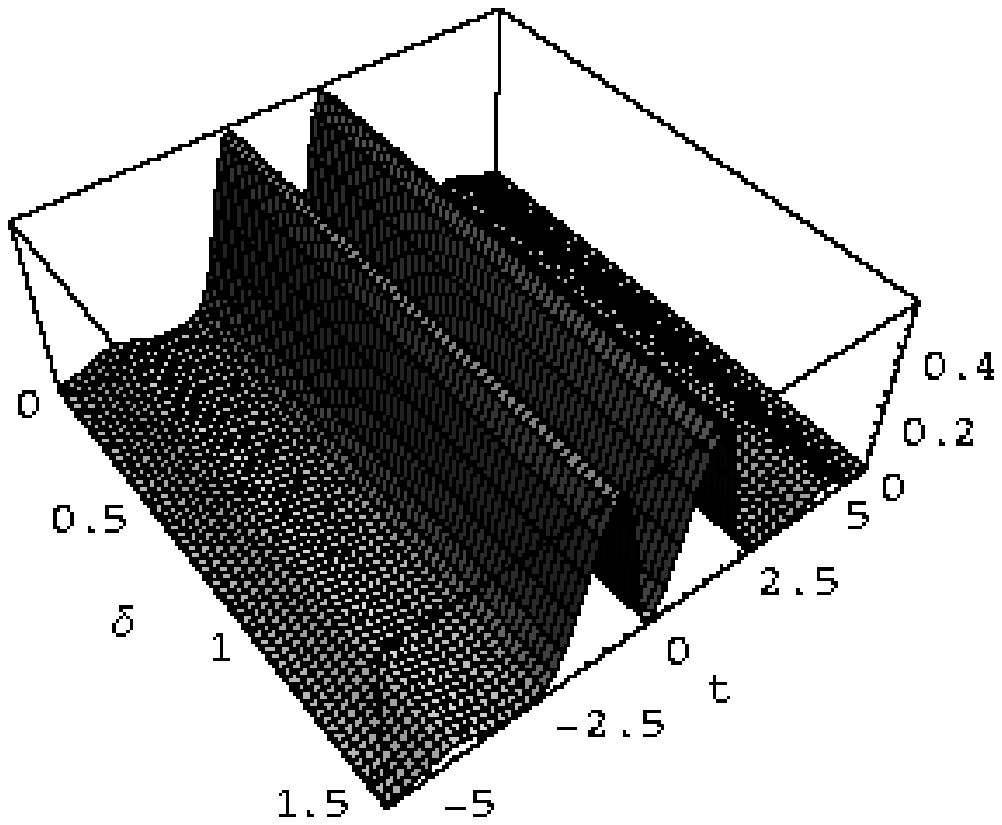}
 \end{tabular}
 \caption{\label{box}  Variation with $t=\tau/ \tau_f$ of $G_+$ (first 2 lines) and $G_-$ (last 2 lines).
 For both $G_+$ and $G_-$, the first line is for $\alpha$ varying between 0
 and $\pi /2$ with $\delta$ fixed at $\delta=\pi /6, \pi/4, \pi/3,
 \pi/2$.
 The second line corresponds to $\delta$ varying between 0 and $\pi /2$
 with $\alpha$ fixed at $\alpha=\pi /6, \pi/4, \pi/3, \pi/2$.}
\end{figure}

\begin{figure}
\includegraphics[height=5cm]{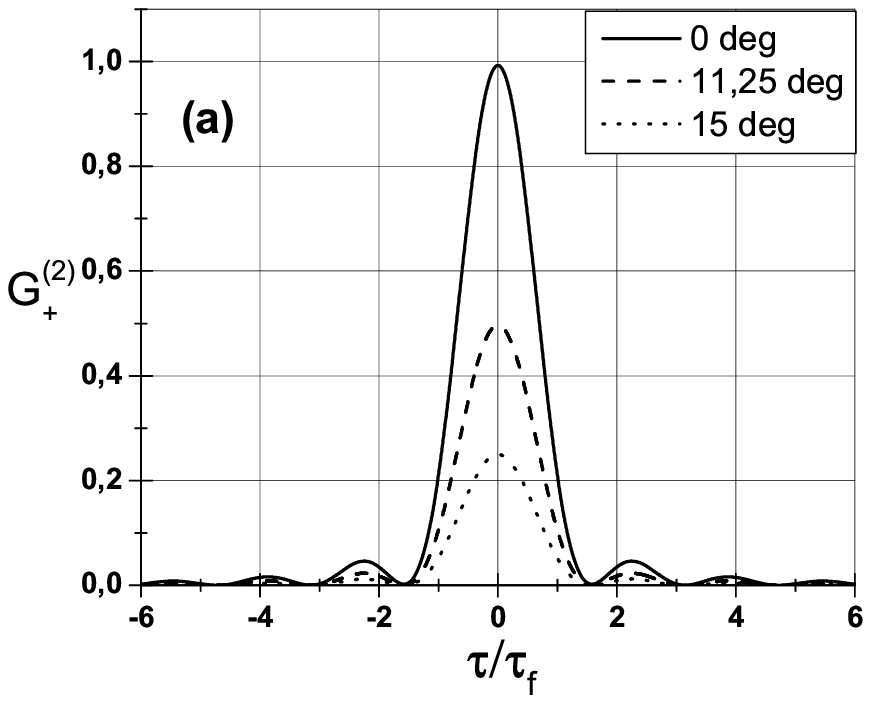}
\includegraphics[height=5cm]{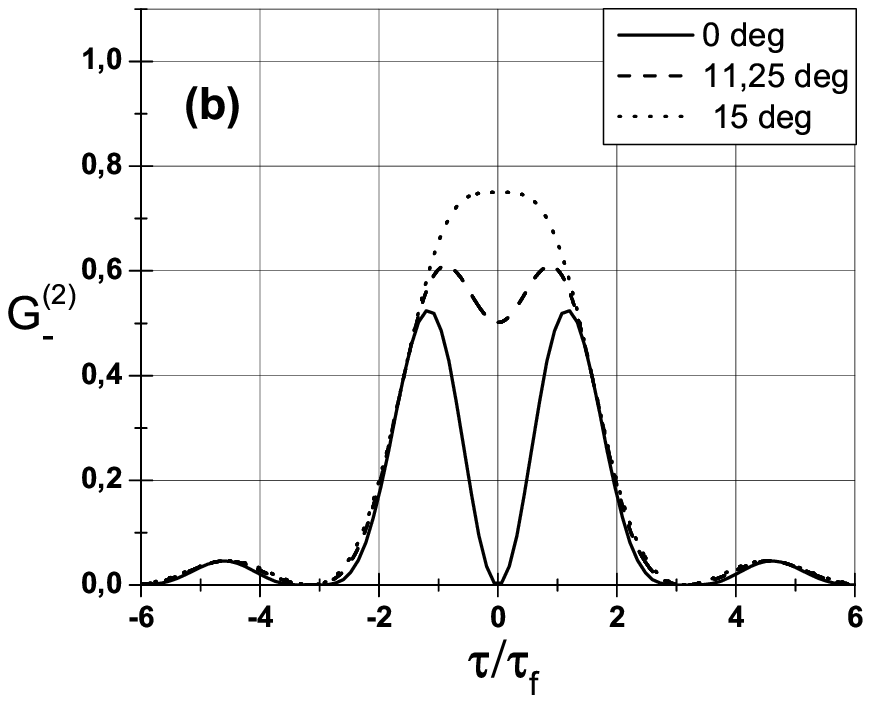}
\caption{The second order intensity correlation function for
different orientations of the HWP acting on type II PDC state in
frequency degenerate regime. $G^{(2)}_{+}$ corresponds to parallel
orientations of polarization filters set to $45^\circ$ (a);
$G^{(2)}_{-}$ corresponds to orthogonal orientations of
polarization filters set to $45^\circ$ and $-45^\circ$ (b).}
\end{figure}

\begin{figure}
\includegraphics[height=5cm]{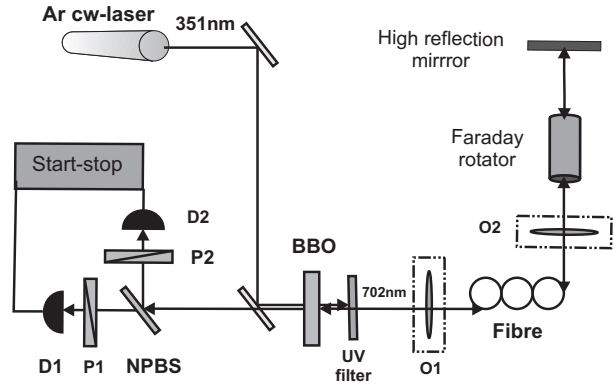} \caption{The experimental
setup. A cw $Ar^+$ laser at 351 nm pumps a type-II BBO crystal; O1,
O2- microscope objectives; compensation for polarization drift in
fibre is performed by free-space Faraday mirror, consisting of
Faraday rotator and high-reflection mirror; NPBS - 50/50
nonpolarizing beamsplitter; P1 and P2- Glan prisms; D1, D2-
avalanche photodiodes. The output of the Start-Stop scheme is
analyzed by a multi-channel analyzer (MCA).}
\end{figure}

\begin{figure}
\includegraphics[height=5cm]{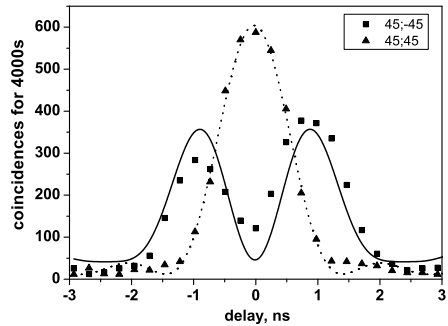}
\caption{Experimental dependence of the coincidence count rate on
the time delay between two photons for two cases: both polarizers
are parallel and set at $45^\circ$ (triangles) and polarizers are
orthogonal and set at $45^\circ$ and $-45^\circ$ (squares). The
dashed and solid curves represent the fit of experimental data for
$45^\circ;45^\circ$ and $45^\circ;-45^\circ$ configurations
respectively.}
\end{figure}


\begin{thebibliography}{25}
\bibitem{twoarms} Here we only consider the case when two photons of the pair go
into different ports of the beamsplitter.
\bibitem{delta} In following calculations we neglet the chromatic dispersion of the
fiber i.e. the optical thickness $\delta$ is considered constant
for all the spectrum of PDC.
\bibitem{visibility} Here we define
the visibility in the traditional way as the difference of
coincidences measured in orthogonal bases divided by their sum.
\end{thebibliography}
\end{document}